\documentclass[
    ,final            
  ]
  {aipproc}
\layoutstyle{6x9}

\begin{document}

\title{Fusion dynamics around the Coulomb barrier}

\author{K. Hagino}{
  address={Yukawa Institute for Theoretical Physics, Kyoto University,
Kyoto 606-8502, Japan}
}

\author{N. Rowley}{
  address={Institut de Recherches Subatomiques, 
UMR7500, IN2P3-CNRS/Universtit\'e Louis Pasteur, BP28, F-67037 
Strasbourg Cedex 2, France}
}

\author{T. Ohtsuki}{
  address={Laboratory of Nuclear Science, Tohoku University, Sendai
982-0826, Japan}
}

\author{M. Dasgupta}{
  address={
Department of Nuclear Physics, Research School of Physical 
Sciences and Engineering, Australian National University, 
Canberra ACT 0200, Australia}
}

\author{J.O. Newton}{
  address={
Department of Nuclear Physics, Research School of Physical 
Sciences and Engineering, Australian National University, 
Canberra ACT 0200, Australia}
}

\author{D.J. Hinde}{
  address={
Department of Nuclear Physics, Research School of Physical 
Sciences and Engineering, Australian National University, 
Canberra ACT 0200, Australia}
}

\begin{abstract}

We perform exact coupled-channels calculations, 
taking into account properly the effects of Coulomb 
coupling and the finite excitation energy of collective excitations in 
the colliding nuclei, 
for three Fm formation reactions, 
$^{37}$Cl + $^{209}$Bi, $^{45}$Sc + $^{197}$Au, 
and $^{59}$Co + $^{181}$Ta. 
For the $^{37}$Cl + $^{209}$Bi and $^{45}$Sc + $^{197}$Au reactions,  
those calculations well reproduce the experimental total fission cross
sections, and a part of the extra-push 
phenomena can be explained in terms of the Coulomb excitations to 
multi-phonon states. 
On the other hand, for the heaviest system, 
the deep-inelastic collisions become much more significant, 
and the fission cross sections are strongly overestimated. 
We also discuss the surprisingly large surface diffuseness parameters 
required to fit recent high-precision fusion data for medium-heavy 
systems, in connection with the fusion supression observed in 
massive systems. 

\end{abstract}

\maketitle


\section{Introduction}

The coupled-channels method has been very successful in reproducing 
experimental cross sections for heavy-ion reactions involving 
{\it medium-heavy} 
nuclei. Particularly, in many systems, 
it simultaneously reproduces 
the subbarrier enhancement of fusion cross sections and the shape of 
the fusion barrier distribution by including 
a few low-lying collective excitations of colliding nuclei and nucleon
transfer channels \cite{DHRS98}. 
It is now a standard theoretical tool to analyse experimental
fusion and quasi-elastic cross sections at energies around the 
Coulomb barrier \cite{HRK99}. 

However, it has not yet been completely understood 
to what extent this method works for the 
fusion of massive systems, where the charge product of the target and
projectile nuclei, $Z_PZ_T$, is typically larger than about 1800. 
For those systems, other 
reaction processes, such as deep-inelastic collision and
quasi-fission, come into play, and the reaction dynamics around the
Coulomb barrier becomes much more complex than for medium-heavy 
systems \cite{R94}. 
In order to calculate fusion cross sections,  
the competition of fusion with these other processes has to be taken 
into account properly \cite{SKA02,ZAIOO02,AWOA99}, 
and the dynamics after the Coulomb barrier is overcome 
becomes very important. 
This is the most difficult problem in the fusion of massive nuclei, and there
are still large ambiguities in theoretical predictions of fusion cross
sections. Because of this reason, one often employs a simplified
coupled-channels treatment for the barrier 
penetration prior to the touching configuration \cite{ZAIOO02,DH00}, 
which is essentially based on the
constant coupling approximation \cite{DLW83} or a variant \cite{MNA92}. 

In this contribution, 
we critically examine the consequences of using such a simplified
coupled-channels framework, 
and point out that the exact treatment for
the Coulomb coupling plays an important role in massive systems. 
We then apply the exact coupled-channels approach to total fission cross
sections for three Fm formation 
reactions \cite{OSA94}, where the charge product 
$Z_PZ_T$ is given by 1411 ($^{37}$Cl + $^{209}$Bi), 
1659 ($^{45}$Sc + $^{197}$Au), 
and 1971 ($^{59}$Co + $^{181}$Ta). 
We will show that the small hindrance of the reduced cross sections for the
second heaviest system compared with the lightest system can be 
understood in terms of the effect of energy loss due to the 
Coulomb excitation, while
the fusion hindrance for the heaviest system exceeds that
effect and an explicit treatment of deep-inelastic collisions is
necessary. 
We also discuss the surface property of the nucleus-nucleus potential, 
where recent high-precision fusion data for medium-heavy systems 
systematically show 
that a surprisingly large diffuseness parameter is required in order
to fit them \cite{NMD01,HDG01,HRD03,NBD03}. 
We argue that this apparant anomaly could originate from the competition 
between fusion and the deep-inelastic processes which occur in fusion
of heavy systems \cite{NBD03}. 

\section{Coupled-channels approach to fusion of massive systems}

\subsection{Effect of Coulomb excitations}

A hot compound nucleus formed by a fusion reaction decays either by emitting
a few particles (mainly neutrons) and gamma rays, or by fission. 
The fusion cross section
is thus a sum of the evaporation residue and the fusion-fission
cross sections. Theoretically, it is computed as 
\begin{equation}
\sigma_{\rm fus}(E)=\frac{\pi}{k^2}\sum_l\,(2l+1)P_{\rm fus}(E,l)
=\frac{\pi}{k^2}\sum_l\,(2l+1)T_{\rm bp}(E,l)\cdot P_{\rm CN}(E,l),
\label{xsection}
\end{equation}
where $T_{\rm bp}(E,l)$ is the barrier passing probability for the 
Coulomb barrier while  
$P_{\rm CN}(E,l)$ is the probability of compound nucleus formation 
after barrier penetration. 

For medium-heavy systems, the compound nucleus is formed
immediately after the Coulomb barrier is overcome, and $P_{\rm CN}$ is 
essentially 1. This justifies the assumption of strong
absorption inside the Coulomb barrier, or equivalently, the incoming
wave boundary condition \cite{HRK99}. The coupled-channels approach
has been successful here. For massive systems, on the other hand, 
$P_{\rm CN}$ significantly deviates from 1, due to competition among
several reaction processes. The fusion cross sections, therefore,
appear to be hindered if one compares fusion cross sections with
those obtained by assuming $P_{\rm CN}=1$. 
Recent experimental data clearly indicate the strong competition 
between compound nucleus formation 
and quasi-fission \cite{BHD01,HDM02,S03}. 

In Eq. (\ref{xsection}), the barrier passing probability 
$T_{\rm bp}$ can in principle be computed by the coupled-channels 
approach. For this, 
although exact coupled-channels codes are available \cite{HRK99}, 
a simplified approach has often been taken, even for fusion of
massive systems \cite{ZAIOO02,DH00}. The simplification is achieved 
by using one or more of the following approximations: i) the linear 
coupling approximation, where the nuclear coupling potential 
is assumed to be linear with respect to an excitation operator 
for intrinsic motions, ii) multi-phonon excitations are neglected, 
and iii) the eigenchannel approximation, where the 
intrinsic excitation energies are treated approximately. 
Since the effective coupling strength is approximately proportional to 
$Z_PZ_T$ for a given value of deformation parameter \cite{HTDHL97}, 
the shortcomings of these approximations become more severe 
for heavier systems. 

The validity of the first and the second
approximations has been examined in detail 
in Refs. \cite{HTDHL97,HTDHL97b,SAC95}. We therefore 
discuss the third point here. The eigenchannel approach is intimately
related to the concept of barrier distribution \cite{DHRS98,DLW83,RSS91}. 
Although this approach is exact when the intrinsic excitation energy 
vanishes, it also works well even with a finite excitation energy as
long as the coupling potential is localized inside the uncoupled 
barrier \cite{HTB97}. 
In realistic cases, however, the coupling potential extends outside
the barrier due to the long range Coulomb interaction. If, prior to 
reaching the barrier, there is
appreciable Coulomb excitation 
to a collective state whose excitation energy is not small, this 
results in a decrease of the
relative energy, leading to the reduction of 
fusion cross sections and a modification of effective eigen-barriers. 
Since the eigenchannel approach treats the 
excitation energy approximately, this effect will be missed in a
calculation. 

\begin{figure}
  \includegraphics[height=.5\textheight,angle=-90]{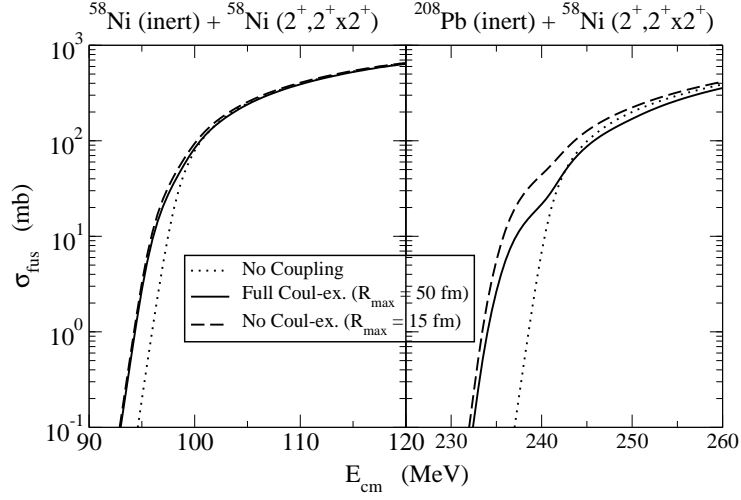}
  \caption{Influence of the Coulomb couplings in 
fusion cross sections for the 
$^{58}$Ni + $^{58}$Ni (left panel) and the $^{58}$Ni + $^{208}$Pb 
(right panel) systems. 
The double quadrupole phonon excitation in the projectile nucleus 
$^{58}$Ni is included in the coupled-channels calculations, while the
target nucleus is treated as inert. The solid line includes the full effect
of the Coulomb excitation, while the dashed line disregards it by
matching the numerical wave functions to the asymptotic wave functions 
at relatively small distance. The fusion
  cross sections without the coupling are denoted by the dotted line. 
}
\end{figure}

In order to demonstrate the effect of Coulomb excitations prior to the
barrier, figure 1 compares fusion cross sections for the 
$^{58}$Ni + $^{58}$Ni system with those for the $^{58}$Ni + $^{208}$Pb 
system. 
We assume that the target nucleus is inert, and include only the
double quadrupole phonon excitation in the projectile 
$^{58}$Ni nucleus. The excitation energy for the single phonon state
is 1.45 MeV, and we assume a simple harmonic oscillator coupling for
the double phonon excitation. 
The solid and dashed lines are obtained by
integrating the coupled-channels equations from inside the Coulomb
barrier to $R_{\rm max}$=50 fm and 15 fm, respectively. 
The latter calculation, therefore, effectively disregards the effect
of Coulomb excitation outside the Coulomb barrier. For a comparison,
we also show the fusion cross sections in the no coupling limit by the
dotted line. One clearly sees that 
the Coulomb coupling considerably alters 
the fusion cross sections when the target is heavy, while the
difference is small for the lighter system. One also notices that the
fusion cross sections are even smaller than the no coupling
calculations for the heavier system at energies above the Coulomb
barrier. Evidently, the Coulomb excitation provides another mechanism
of fusion inhibition in massive systems. 

\subsection{Total fission cross sections for Fm formation reactions}

\begin{figure}
  \includegraphics[height=.5\textheight,angle=-90]{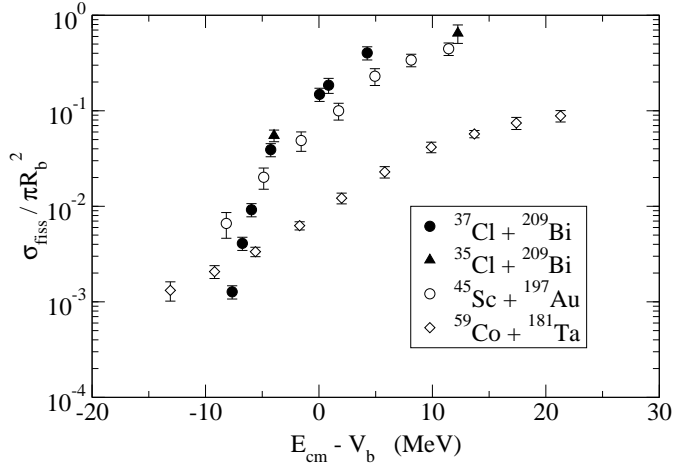}
  \caption{Reduced total fission cross sections for four reactions
  forming Fm, 
$^{35,37}$Cl + $^{209}$Bi (the filled circles and triangles), 
$^{45}$Sc + $^{197}$Au (the open circles), 
and $^{59}$Co + $^{181}$Ta (the open diamonds), as a function of
energy measured from the Coulomb barrier height.}
\end{figure}

Let us now 
discuss fission cross
sections for $Z$=100 (Fm) formation reactions measured by Ohtsuki {\it
et al.} \cite{OSA94}. The experimental cross sections were obtained by 
selecting the fission events in the TOF vs $\Delta E$ plot, and thus 
contain both the fusion-fission and the quasi-fission (if any) 
cross sections. 
For the heaviest system, the fission events were not well separated 
from the deep-inelastic collision (DIC) events. 
For this system, the fission cross sections were estimated by choosing 
the same mass region of fission fragments as that observed in the 
$^{35,37}$Cl + $^{209}$Bi reaction. 
Therefore, the experimental fission cross sections might be 
underestimated for the heaviest system 
if the fission events extend towards the DIC region. 
Fig. 2 shows the observed 
reduced fission cross sections as a 
function of the 
difference between the 
center of mass energy and the average barrier energy 
for four systems, 
$^{35,37}$Cl + $^{209}$Bi ($Z_PZ_T$=1411), $^{45}$Sc + $^{197}$Au 
($Z_PZ_T$=1659), 
and $^{59}$Co + $^{181}$Ta ($Z_PZ_T$=1971). Notice that 
the reduced cross sections for the second heaviest system 
($^{45}$Sc + $^{197}$Au) are somewhat smaller than those for the
lightest system ($^{35,37}$Cl + $^{209}$Bi), and the cross sections
for the heaviest system ($^{59}$Co + $^{181}$Ta) are substantially 
hindered compared with the other two systems. 

\begin{figure}
  \includegraphics[height=.5\textheight]{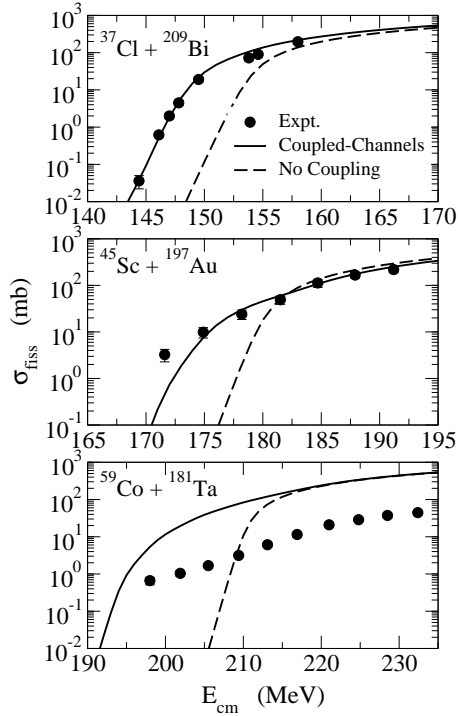}
  \caption{Comparison of the coupled-channels calculations with the
  experimental data for total fission cross sections for the 
Fm formation reactions. The solid line is the result of the
  coupled-channels calculations, while the dashed line is obtained 
without any coupling. The experimental data are denoted by the filled 
  circles.}
\end{figure}

In order to investigate whether these features can be understood in terms of
the Coulomb excitation discussed in the previous subsection, we perform 
exact coupled-channels calculations. 
For this purpose, we use an extended version of the computer code CCFULL
\cite{HRK99}, where the effect of finite ground state spin of
colliding nuclei 
is incorporated within the isocentrifugal approximation. 
For the $^{37}$Cl + $^{209}$Bi system, 
we include three vibrational states, 5/2$^+$ (3.09 MeV), 
7/2$^-$ (3.1 MeV), and 9/2$^-$ (4.01 MeV) in 
$^{37}$Cl as well as double octupole vibrations in $^{209}$Bi. 
The deformation parameters are estimated from the experimental B(E2)
and B(E3) values. We introduce a single effective channel for the
seven octupole states which have a character of 
$h_{9/2}\otimes$ $^{208}$Pb(3$^-$) in $^{209}$Bi, and consider a
harmonic oscillator coupling for the double phonon state. 
For the $^{45}$Sc + $^{197}$Au system, we include 5 quadrupole 
states which have a character of 
$f_{7/2}\otimes$ $^{44}$Ca(2$^+$) in $^{45}$Sc [3/2$^-$ (0.38 MeV), 
5/2$^-$ (0.72 MeV), 11/2$^-$ (1.24 MeV), 7/2$^-$ (1.41 MeV), 
and 9/2$^-$ (1.66 MeV)], and treat the target nucleus 
$^{197}$Au as a classical rotor with $\beta_2=-0.13$ and 
$\beta_4=-0.03$. 
For  $^{45}$Sc, 
since the excitation energies for the quadrupole states are not close to each
other, we do not introduce an effective single channel,  
but treat them exactly. 
For the $^{59}$Co + $^{181}$Ta system, we include 5 quadrupole 
states which have a character of 
$(f_{7/2})^{-1}\otimes$ $^{60}$Ni(2$^+$) in $^{59}$Co [3/2$^-$ (1.1 MeV), 
9/2$^-$ (1.19 MeV), 11/2$^-$ (1.46 MeV), 5/2$^-$ (1.48 MeV), 
and 7/2$^-$ (1.74 MeV)], and treat the target nucleus 
$^{181}$Ta as a classical rotor with $\beta_2=0.354$ and 
$\beta_4=-0.05$. 
More details of the calculations will be given elsewhere
\cite{OIN03}. 

The results of those calculations are shown in Fig. 3. 
One finds that the coupled-channels calculations well reproduce the
experimental data for the two lightest systems, 
$^{37}$Cl + $^{209}$Bi and $^{45}$Sc + $^{197}$Au. Especially, the
reduction of cross sections in the latter system is reproduced 
nicely. As we discussed in the previous subsection, this small
hindrance of cross sections is caused by the strong Coulomb coupling 
to the collective states outside the Coulomb barrier. 
In contrast, the coupled-channels calculation 
considerably overestimates fission
cross sections for the heaviest system, $^{59}$Co + $^{181}$Ta. 
We will discuss this point in the next subsection. 

\subsection{Role of deep-inelastic collision}

\begin{figure}
  \includegraphics[height=.4\textheight,angle=-90]{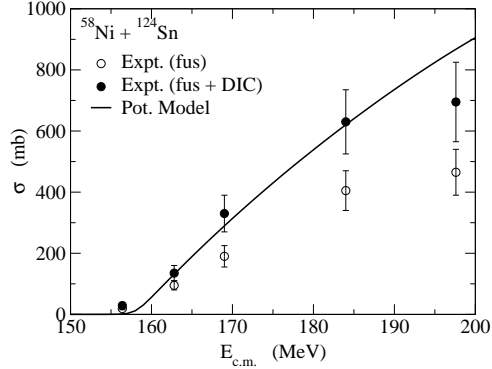}
  \caption{Fusion cross sections (the open circles) and the sum of
  fusion and deep inelastic collision cross sections (the filled 
  circles) for the $^{58}$Ni + $^{124}$Sn reaction. 
The solid line is a prediction of the barrier penetration model 
with a global nucleus-nucleus potential.
The experimental data are taken from Ref. \cite{W87}.}
\end{figure}

In his review article, Reisdorf argued \cite{R94} that cross sections
for the sum of fusion and other damped reactions may be interpreted 
as the total barrier passing cross sections. Indeed, he found that 
the sum of fusion and deep-inelastic collision (DIC) cross sections 
for the $^{58}$Ni + $^{124}$Sn reaction could be well reproduced by the
standard potential model (see Fig. 4). More recently, Esbensen {\it et
al.} followed a similar idea and reproduced the experimental cross
sections for the sum of fusion and DIC reactions for the same system 
with the coupled-channels approach \cite{EJR98}. 
In the semiclassical picture, deep inelastic collisions correspond 
to those trajectories which overcome the barrier in the entrance
channel but eventually escape after appreciable interaction with
the target. 
All of those considerations immediately lead to the idea that the
total barrier passing cross sections, which the coupled-channels
approach yields, may have to be compared with a sum of fusion, 
quasi-fission, and DIC, {\it i.e.,} 
\begin{equation}
\sigma_{\rm bp}(E)=\sigma_{\rm fus}(E)
+\sigma_{\rm qf}(E)+\sigma_{\rm DIC}(E). 
\end{equation}
The large reduction of total fission cross sections for the 
$^{59}$Co + $^{181}$Ta system, therefore, may be attributed to
the competition between fusion, quasi-fission and DIC reactions. 

The present framework of the coupled-channels method could be used to obtain
inclusive cross sections of fusion and DIC reactions, but it would be
difficult to obtain them separately. 
One possible way to isolate fusion from DIC theoretically is to introduce 
angular momentum truncation in
Eq. (\ref{xsection}) and exclude higher partial wave contributions, as
was done by Zagrebaev {\it et al.} \cite{ZAIOO02}. 
However, a large ambiguity exists in this prescription, since 
there is no clear and unique way to introduce the angular momentum truncation, 
especially at 
energies below the barrier \cite{DP89}. 
In recent publications, 
Abe {\it et al.} combined the surface friction model with the Langevin
approach in order to take into account the competition between fusion
and DIC in the approaching phase for a formation reaction 
of superheavy elements \cite{SKA02}.  
This approach may be promising, but is essentially classical. 
Computation of fusion cross
sections with a quantum mechanical model under the influence of DIC
process is still an open problem. 

\section{Surface diffuseness anomaly in fusion potentials} 

\begin{figure}
  \includegraphics[height=.4\textheight,angle=-90]{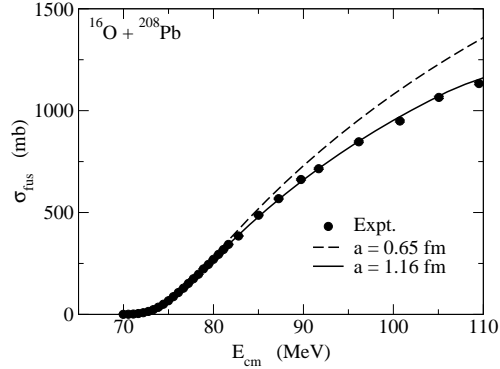}
  \caption{
Comparison of coupled-channels calculations 
with the experimental data for 
the $^{16}$O + $^{208}$Pb fusion reaction. 
The double octupole phonon as well as the single $5^-$ phonon
  excitations are included. 
The solid and the dashed lines are obtained by setting the surface 
diffuseness parameter
of Woods-Saxon potential to be $a$=1.16 and $a$= 0.65 fm,
  respectively. The experimental data are taken from Ref. \cite{M99}.}
\end{figure}

Let us now discuss the second subject, that is the surface 
property of the nucleus-nucleus potential and its anomaly, recently
recognised in fusion reactions for medium-heavy systems. 
For scattering processes, it seems well accepted that
the surface diffuseness parameter $a$ should be around 0.63 fm
\cite{BW91} 
if the nuclear potential is parametrised by a Woods-Saxon (WS) form. 
In marked contrast, recent high-precision 
fusion data suggest that a much larger 
diffuseness, between 0.8 and 1.4 fm, is needed 
to fit the data \cite{NMD01,HDG01}. This is not just for particular 
systems but seems to be a rather general 
result \cite{NBD03}. 

We illustrate this problem in Fig. 5 by comparing experimental data 
for the $^{16}$O + $^{208}$Pb fusion reaction with 
various coupled-channels calculations with the WS potential. 
We include the double octupole phonon and the single 5$^-$ phonon 
excitations in $^{208}$Pb. 
At energies well above the barrier, 
where the fusion cross sections $\sigma_{\rm fus}$ is relatively
insensitive to the couplings, a WS potential with a diffuseness 
$a$ =0.65 fm significantly 
overestimates fusion cross sections (dashed line). 
Changing the depth and
radius parameters in the WS potential is not helpful,
since it merely leads to an energy shift in the calculated fusion 
cross sections 
without significantly changing the energy dependence. 
On the other hand, a potential with $a$=1.16 fm (full line) fits the
data well. A similar problem also exists at energies below the 
barrier \cite{HDG01,HRD03}, but we do not discuss it in this
contribution. 

\begin{figure}
  \includegraphics[height=.4\textheight]{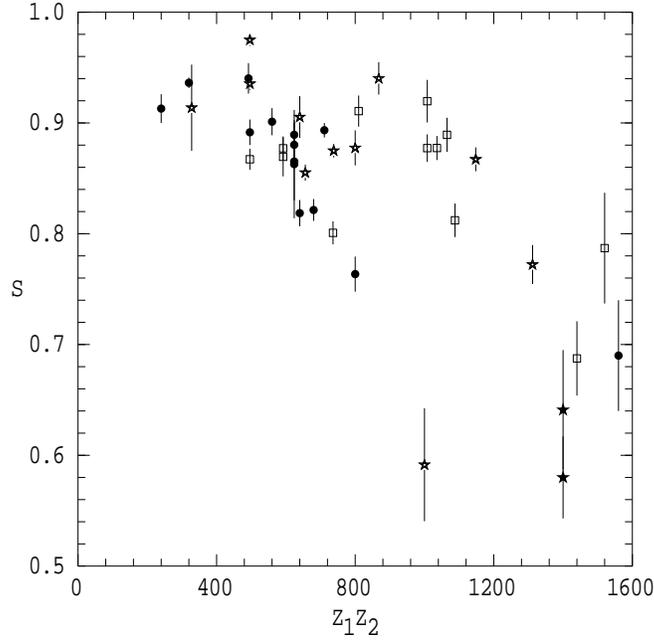}
  \caption{
Calculated suppression factors $S$ for fusion cross sections with
  respect to the prediction of potential model with $a\approx$0.63 fm. 
The large filled stars at $Z_1Z_2$=1400 refer to the reactions 
$^{58}$Ni + $^{112,124}$Sn, where the deep-inelastic collisions have
  been observed experimentally.}
\end{figure}

Up to now, several possible reasons for this anomaly have been
considered. These include 
the departure of the nuclear potential from the WS form
\cite{NMD01,HDG01}, dissipation effects \cite{HDG01}, and unrecognised  
systematic errors in experimental data \cite{NBD03}, 
but none of them is conclusive yet. 
We would like to propose here another possible effect, 
that is, the 
competition between fusion and deep inelastic collision \cite{NBD03}. 
This is motivated by an 
apparent similarity between fusion inhibition in massive
systems discussed in the previous section (see fig. 4) and the overestimate of
fusion cross sections shown in fig. 5. 
To this end, let us introduce a suppression factor 
$S$ which is defined as a ratio between the experimental fusion cross
sections and the prediction of the potential model with the standard 
value for the diffuseness parameter, $a\approx$0.63 fm. 
Provided that $S$ is independent of bombarding energy,
it can be stated that the fusion cross sections are hindered by a
factor $S$ for whatever reason. 

The values of $S$ obtained for a large number of systems are shown in
fig. 6 \cite{NBD03}. 
Also shown are two points (large filled stars) at $Z_1Z_2$ = 1400 
derived from the data for the 
$^{58}$Ni + $^{112,124}$Sn reactions at energies around the fusion
barrier. For these systems, as we indicate in Fig. 4, a very 
significant contribution from deep-inelastic scattering has been
observed experimentally even at energies below the barrier
\cite{W87}. In this case, the value of $S$ has been taken as
$\sigma_{\rm fus}$ divided by the sum of 
$\sigma_{\rm fus}$ and $\sigma_{\rm DIC}$. 
These two points do not seem inconsistent with the other points 
with large $Z_1Z_2$ in Fig. 6, which are 
derived from the fusion data only. 
Strongly damped reactions similar to DIC are reported to occur in
lighter systems, for example, 
$^{32}$S + $^{64}$Ni \cite{RAA89} ($Z_1Z_2$=448) at energies well
above the barrier. It would be interesting to see if they also occur
at energies closer to the barrier. If they do, the values of $S$ would
be reduced below unity and might result in at least a partial
explanation of the experimental values for $a$ being much larger than 
the standard value, 0.63 fm. 

\section{Summary}

Extensive efforts have been made both experimentally and
theoretically to understand fusion of massive systems, especially for 
synthesis of superheavy elements (SHE), but the reaction mechanism 
has not yet been completely understood. 
Here, we have performed coupled-channels calculations and pointed
out that the exact treatment of Coulomb excitation becomes important 
for massive systems. We have applied the coupled-channels framework to
three Fm formation reactions, and have shown that Coulomb
excitation indeed provides an important mechanism of inhibition of
fusion cross sections for transitional systems  (``pre-SHE'' systems) 
between medium-heavy and SHE regions. 
For SHE systems, we have argued that the deep-inelastic collision 
needs to be taken into account explicitly in theoretical models. 
However, it is still an open problem of how to incorporate the competition
between fusion and DIC quantum mechanically, and further developments 
will be required. 
We have also discussed the large surface diffuseness problem observed
in the recent high precision measurements of fusion cross sections 
for medium-heavy systems in connection with the fusion hindrance in
massive systems. We have argued that the competition between fusion 
and deep-inelastic collision may provide a promising avenue to explain
this anomaly. In this connection, it would be interesting to study, both
experimentally and theoretically, light systems near to the fusion
barrier to see whether fusion was inhibited by the presence of DIC. 
Especially, theoretical calculations involving friction for light
systems would be of great interest, since experimental measurements 
are likely to be difficult when the fusion suppression factor 
appoaches unity. 



\end{document}